\documentclass[twocolumn,apjl]{emulateapj}
\usepackage{color}
\usepackage{ulem,cancel,amsmath}

\def\mbf#1{\mbox{\boldmath ${#1}$}}
\def\Alfven{Alfv\'{e}n~}
\def\Alfvenic{Alfv\'{e}nic~}

\begin{document}

\title{Atmospheric Escape by Magnetically Driven Wind from Gaseous Planets II --Effects of Magnetic Diffusion--}

\author{Yuki A. Tanaka$^{1}$, Takeru K. Suzuki$^{1}$ 
\& Shu-ichiro Inutsuka$^{1}$}
\email{tanaka.yuki@b.mbox.nagoya-u.ac.jp}
\altaffiltext{1}{Department of Physics, Nagoya University,
Nagoya, Aichi 464-8602, Japan}

\begin{abstract}
We investigate roles of \Alfvenic waves in the weakly-ionized atmosphere of hot Jupiters by carrying out non-ideal magnetohydrodynamic (MHD) simulations with Ohmic diffusion in one-dimensional magnetic flux tubes.
Turbulence at the surface excites \Alfven waves and they propagate upward to drive hot ($\approx 10^4$ K) outflows.
The magnetic diffusion plays an important role in the dissipation of the \Alfvenic waves in the weakly ionized atmosphere of hot Jupiters. 
The mass-loss rate of the spontaneously driven planetary wind is considerably reduced, in comparison with that obtained from ideal MHD simulations because the \Alfvenic waves are severely damped at low altitudes in the atmosphere, whereas the wave heating is still important in the heating of the upper atmosphere.
Dependence on the surface temperature, planetary radius, and velocity dispersion at the surface is also investigated.
We find an inversion phenomenon of the transmitted wave energy flux; the energy flux carried by \Alfven waves in the upper atmosphere has a nonmonotonic correlation with the input energy flux from the surface in a certain range of the surface temperature because the resistivity is determined by the global physical properties of the atmosphere in a complicated manner.
We also point out that the heating and mass loss are expected only in limited zones if the open magnetic field is confined in the limited regions.
\end{abstract}
\keywords{planets and satellites: atmospheres --- magnetohydrodynamics (MHD) --- planets and satellites: magnetic fields --- planets and satellites: gaseous planets}


\section{Introduction}\label{introduction}


Recently, a large number of exoplanets have been found by various detection methods, and it have been learned that the exoplanets are rich in diversity.
Some of them have very small semi-major axes and masses that are similar to the Jupiter, and they are known as hot Jupiters.

Transit observations are a very useful method to know not only the radius of planets and their orbital period, but also information about their atmosphere.
For example, transit observations in the UV band provide interesting physical properties of the upper atmosphere.
Absorption in the H Ly$\alpha$ have been confirmed in some hot Jupiters, such as HD 209458b \citep{vid03}, HD 189733b \citep{lec10}, and 55 Cnc b \citep{ehr12}.
The transit depth corresponds to the apparent radius of planets.
While the transit depth in optical wavelength at HD 209458b is $\sim$ 1.5\%, that corresponds to $\sim$ 1.38 $R_{J}$, the transit depth in the H Ly$\alpha$ line is $\sim$ 15\%, that corresponds to $\sim$ 4.3 $R_{J}$ \citep{vid03}.
Similar phenomena are reported at HD 189733b \citep{lec10}, and a hot Neptune, GJ 436b \citep{kul14}.
These observations suggest that the existence of the extended exosphere that is filled up with the high-temperature atomic hydrogen gas, and mass loss takes place.
Heavier elements such as C, O, Si, and Mg are also detected in the UV band in the escaping atmosphere \citep[e.g.,][]{vid04, lin10, bb13}.

Although the detailed mechanism of the strong mass loss from hot Jupiters is still unclear, various theoretical models have been developed.
A leading example of them is that energy-limited escape by X-ray and extreme ultraviolet (XUV) from a central star, so-called XUV-driven escape \citep{lam03}.
The XUV-driven escape have been studied by various ways; for example, including chemical reactions \citep{yel04, yel06}, considering radiation pressure on the escaping atmosphere and the charge exchange between neutral particles in the escaping atmosphere and protons in the hot stellar wind \citep{hol08, eke10, bl13, tc13}.
Additionally, it is suggested that XUV-driven escape can occur not only in hot Jupiters but also in cooler planets.
\cite{cha15} demonstrated that hydrodynamic escape driven by XUV can take place at the gaseous planets that orbit beyond 1 AU in some cases if the central stars are active in the X-ray.

Furthermore, physical properties and phenomena related to magnetic fields of exoplanets have been investigated in more detail by both theoretical and observational studies.
For the theoretical works, for example, \cite{ada11, oa14} suggested outflow from the upper atmosphere will be magnetically-controlled, thus the outflow becomes asymmetric and a mass-loss rate can be smaller about an order of magnitude than the value without the magnetic fields, that is based on the assumption of the spherical symmetric wind.
Also, models of the upper atmosphere with the planetary magnetic fields are proposed and the effects on transit depth and the  loss of the angular momentum by the planetary magnetic fields were discussed \citep{tra11, tra14}.
Recently, magnetic activities of exoplanets have been observed, and several works suggested that the strength of the magnetic fields of hot Jupiters seems to be smaller than that of Jupiter \citep[e.g.,][]{kis14}

We have also investigated magnetically-driven mass loss from gaseous planets by extending MHD simulations for winds from stars with surface convection, e.g. solar wind \citep{tan14}. \Alfven waves, which are excited by turbulence at the surface, transport the energy upward.
They heat up the upper atomosphere and drive outflows. 
However, in this paper we assumed the ideal MHD approximation, which should be modified to take into account effects of the weak ionization in the planet atmopshere.
In this paper, we study non-ideal MHD effects by considering the magnetic diffusion in partially-ionized atmospheres.

We explain our concept of the simulations and numerical method, and several formulae for analysis in Section \ref{model_description}.
Then we show results on the structure of atmopsheres and outflows by inspecting roles of the magnetic resistivity in the propagation and dissipation of \Alfven waves in Section \ref{results}.
In Section \ref{discussion}, we discuss effects of Ohmic dissipation in the atmosphere and effects of geometry of the magnetic flux tubes.
In Section \ref{conclusion}, we summarize the results and refer to our future works.


\section{Model Description}\label{model_description}



\subsection{Numerical Method}\label{numerical_method}


In this work we use a numerical simulation code that is originally developed for the calculation of the acceleration of solar wind \citep{si05,si06}.
At the surface of stars, various types of magnetic waves are excited \citep{ms12,ms14}.
The excited waves propagate upward from the surface, then heat up the coronal region and drive the stellar wind by various dissipation processes of the waves \citep{gol78,hp83,ter86, ks99,mat99}.
It is thought that the \Alfven wave is especially important for the transfer of the energy.
This mechanism of the energy transport from the surface to the upper atmosphere can be applied for objects that have their own magnetic field.
Although details are uncertain, hot Jupiters are expected to have magnetic fields, so this mechanism is applicable to hot Jupiters.
In our previous research, we have extended the simulation code to calculate the atmospheric escape from gaseous planets, especially from hot Jupiters \citep{tan14}.
In addition, this mechanism is applied for the calculation of the structure of the atmosphere of brown dwarfs, and it is implied that the heating by the dissipation of MHD waves is also important for the brown dwarfs' atmosphere \citep{sor14}.

As for our MHD simulations we basically follow \citep{tan14}, except that we consider magnetic diffusion; we solve time-dependent equations and treat the propagation and dissipation of the MHD waves in a super-radially open magnetic flux tube in the atmosphere.
We adopt the same functional form for the super-radial expansion factor, $f(r)$, as in \citep{tan14}, which was originally introduced by \citep{kh76}.
\begin{equation}
\label{eq:fr}
f\!\left(r\right)=\frac{e^{\frac{r-r_{0}-h_{1}}{h_{1}}}+f_{0}-\left(1-f_{0}\right)/e}{e^{\frac{r-r_{0}-h_{1}}{h_{1}}}+1},
\end{equation}
where $h_{1}$ is the typical height of closed magnetic loops of, and the subscript 0 means the surface.

The effects of resistivity explicitly appear in the energy equation and induction equation:
$$
\rho\frac{d}{dt}\!\left(e+\frac{v^{2}}{2}+\frac{B^{2}}{8\pi\rho}-\frac{GM_{\star}}{r}\right)
+
\frac{1}{r^{2}f}\frac{\partial}{\partial r}\!\left\{r^{2}f\left[\left(p+\frac{B^{2}}{8\pi}\right)v_{r}\right.\right.
$$
$$
\left.\left. -\frac{B_{r}}{4\pi}\left(\mbf{B\cdot v}\right)
-
\frac{\eta}{4\pi}\frac{B_{\perp}}{r\sqrt{\mathstrut f}}\frac{\partial}{\partial r}\!\left(r\sqrt{\mathstrut f}B_{\perp}\right)\right]\right\}
$$
\begin{equation}
\label{eq:ene}
+\frac{1}{r^{2}f}\frac{\partial}{\partial r}\!\left(r^{2}fF_{c}\right)+q_{R}=0,
\end{equation}
\begin{equation}
\frac{\partial B_{\perp}}{\partial t}
=
\frac{1}{r\sqrt{\mathstrut f}}\frac{\partial}{\partial r}\!
\left[r\sqrt{\mathstrut f}\left(v_{\perp}B_{r}-v_{r}B_{\perp}\right)
+
\eta\frac{\partial}{\partial r}\!\left(r\sqrt{\mathstrut f}B_{\perp}\right)\right],
\label{eq:ind}
\end{equation}
where $\rho$, $\mbf{v}$, $p$, $e$, and $\mbf{B}$ are the density, velocity, pressure, specific energy, and magnetic field strength, respectively.
The subscripts $r$ and $\perp$ denote the radial and perpendicular components; $d/dt$ and $\partial/\partial t$ denote the Lagrangian and Eulerian derivatives.
$G$ and $M_{\star}$ denote the gravitational constant and the mass of a central object, and $\eta$ is the magnetic resistivity.
$F_{c}$ and $q_{R}$ are the thermal conductive flux, and radiative cooling \citep[][see \cite{tan14} for detail]{lm90,sd93,aa89a}.
In these equations, the effect of the super-radial expansion of the magnetic flux tubes appears as the factor $r\sqrt{\mathstrut f }$.
Note that the term with $\eta$ in the energy equation (\ref{eq:ene}) indicates Joule (Ohmic) heating and the term with $\eta$ in the induction equation (\ref{eq:ind}) denotes diffusion of magnetic field lines. 

The most dominant origin of the magnetic resistivity in the atmosphere of hot Jupiters is the collision between electrons and neutral particles.
In such circumstances the magnetic resistivity $\eta$ depends on gas temperature and an ionization degree as follows;
\begin{equation}
\label{eq:eta}
\eta\approx234\frac{\sqrt{\mathstrut T\left(K\right)}}{x_{e}}\left({\rm cm^{2}\,s^{-1}}\right),
\end{equation}
where $x_{e}=n_{e}/n$ is the ionization degree \citep[e.g.,][]{bb94}.
$n_{e}$ is number density of electrons, and $n$ is that of neutral and ionized hydrogen.
If the second term on the right-hand side of Equation (\ref{eq:ind}) is dominant over the first term, that is, the magnetic resistivity is large, the magnetic waves will diffuse out.
Since the gas temperature and ionization degree will be low in the atmosphere of planets, the magnetic resistivity might not be negligible.
Therefore, the resistive MHD calculation is needed to evaluate the non-ideal effects on the structure of the atmosphere and the mass loss.

Ionization degrees are determined by a method originally developed for Betelgeuse by \cite{ha84} \citep[see also][]{har09}.
We adopt the solar abundance gas \citep[][]{ag89}\footnote{Note: the ``solar composition'' is updated recently by \cite{asp09}}, and calculate the ionization and recombination of H, C, Na, Mg, Al, Si, S, Ca, Fe, and K in the gas phase. 
Since the typical surface temperature of hot Jupiters is $\sim$ 1000 K and it is low to ionize hydrogen, heavy elements such as Na and K are the dominant ionizing sources in the lower atmosphere.

The inner boundary of the simulations is set at the photosphere in the solar and stellar wind calculations \citep{si05,si06,su07,su13}.
In this paper, we set the inner boundary at the position where $p_{0}=10^{5}\,{\rm dyn\,cm^{-2}}(=0.1{\rm bar})$.
We fix the temperature at the inner boundary, and it is treated as a parameter.
The density at the inner boundary is determined by the given temperature and $p_{0}$.
The outer boundary radius is set to 360 planetary radii, and the number of the grids is 6000.
In the most inner region, $dr$ corresponds to 0.0001 planetary radii, and $dr$ increases gradually as the distance from the surface.

We set the planet mass to be the Jupiter mass throughout this paper.
We inject velocity perturbations at the inner boundary, which excite upward propagating \Alfven waves, and their value is treated as a parameter.
In most cases, amplitude of the injected perturbations is 20 \% of the sound speed at the surface.
In the case of the sun, the velocity dispersion at the surface is caused by surface convection.
On the other hand, several previous works have suggested that the convective-radiative boundary lies deep region of the atmosphere\citep[e.g.,][]{bur03}.
However, three-dimensional calculations of the atmospheric circulation of hot Jupiters suggest the existence of supersonic equatorial flow\citep[e.g.,][]{sg02}.
Therefore, it is expected that turbulence can be created at the surface by the strong wind.
We assume a broadband spectrum of the perturbations in proportion to $1/\nu$, where $\nu$ is frequency of perturbations. 
To solve the MHD equations, we adopt a second-order Godunov scheme with the Method of Characteristics \citep{san99}.


\subsection{Useful Formulae}\label{useful_formulae}


We here summarize several useful formula to analyze and evaluate simulation results.


\subsubsection{Scale Height}\label{scale_height}


We treat the surface temperature $T_{0}$, the radius of a planet $R_{p}$, and the velocity dispersion $\delta v$ at the surface as parameters.
We discuss the dependence of the atmospheric structure on these parameters.
To evaluate them, we introduce the scale height of the atmosphere.
The equation of hydrostatic equilibrium of the atmosphere is
\begin{eqnarray}
\frac{1}{\rho}\frac{dp}{dr}+\frac{GM_{p}}{r^{2}}=0.
\end{eqnarray}
$M_{p}$ is the mass of a planet.
If we assume isothermal condition, the density profile can be expressed as follows;
\begin{eqnarray}
\rho&=&\rho_{0}\exp\!\left[-\frac{r-R_{p}}{H_{0}}\frac{R_{p}}{r}\right]\nonumber\\
&=&\rho_{0}\exp\!\left[-\frac{mg}{k_{\rm B}T_{0}}\frac{\left(r-R_{p}\right)R_{p}}{r}\right],\label{densitystructure}
\end{eqnarray}
where $\rho_{0}$ is the density at the surface, $r=R_{p}$ and $H_{0}$ is the pressure scale height of the atmosphere that is denoted as
\begin{eqnarray}
H_{0}=\frac{k_{\rm B}T_{0}}{mg},\label{scaleheight}
\end{eqnarray}
where $k_{\rm B}$ is the Boltzmann constant, $m$ is mean molecular weight, and $g$ is the gravitational acceleration at the planetary surface.


\subsubsection{Poynting Flux}\label{Poynting_flux}


Behavior of \Alfvenic wave in the atmosphere, such as propagation, reflection, and dissipation, is a key to understand the atmospheric structure.
In the ideal MHD simulations the dissipation of the \Alfvenic waves is done by the nonlinear mode conversion.
The fluctuations of the magnetic pressure associated with the \Alfvenic waves excite compressive waves.
These waves are eventually steepen to shocklets, which finally dissipate and heat the surrounding gas \citep{tan14}. 
This process has been examined in terms of the heating of the solar corona and the acceleration of the solar wind \citep{hol82,ks99,si05}.
Recently, a radio observation by JAXA's spacecraft {\it Akatsuki} actually reveals radial distribution of compressive waves in the solar corona, which supports this scenario \citep{miy14}.

In the resistive MHD simulations, the magnetic diffusion also contributes to the dissipation of MHD waves.
When $T_{0}$ is lower, the dissipation of MHD waves in the low altitude where the magnetic resistivity is high affects the structure of the atmosphere.
Therefore, an energetics argument of propagating MHD waves in the resistive MHD condition is important to understand the properties of the atmospheric structure and mass loss.

To evaluate the propagation, reflection, and dissipation of the \Alfvenic wave energy in the atmosphere, we summarize the Poynting flux carried by \Alfven waves below.
The net Poynting flux arising from magnetic tension, which represents the energy flux of \Alfven waves measured from the comoving frame, can be written as follows; 
\begin{eqnarray}
F_{\rm P}= -B_{r}\frac{v_{\perp}B_{\perp}}{4\pi}.\label{Pflux}
\end{eqnarray}
The Poynting flux can be divided into an outward (parallel with $B_r$) part and an inward (antiparallel with $B_r$) part \citep{jac77,cra07,su13}.
In order to evaluate the inward and outward flux, we introduce Els\"asser variables,
\begin{eqnarray}
z_{\pm}=v_{\perp}\mp\frac{B_{\perp}}{\sqrt{\mathstrut 4\pi\rho}}.
\end{eqnarray}
By using the Els\"asser variables, the net Poynting flux is
\begin{eqnarray}
F_{\rm P}=\frac{1}{4}\rho\left(z_{+}^{2}-z_{-}^{2}\right)v_{\rm A}.\label{Pflux2}
\end{eqnarray}
In this equation, $\rho z_{+}^{2}v_{\rm A}/4$ and $\rho z_{-}^{2}v_{\rm A}/4$ correspond to the outward and inward components of the Poynting flux, respectively.
Although in our simulations we inject the outward component of \Alfvenic waves from the surface, the waves are eventually reflected inwardly.


\section{Results}\label{results}



\subsection{Role of Ohmic Dissipation}\label{role_of_resistive_dissipation}


In resistive MHD simulations, the propagation and dissipation of MHD waves are affected by magnetic diffusivity, and accordingly the atmospheric structure would be modified. 
We compare the ideal MHD and resistive MHD calculations, and investigate how the magnetic diffusion affects the structure of planetary atmospheres and outflows.

\begin{figure}[h]
\includegraphics[]{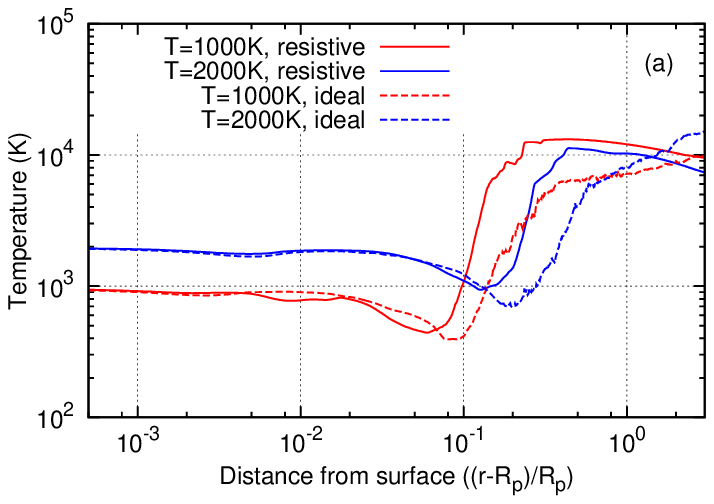}
\includegraphics[]{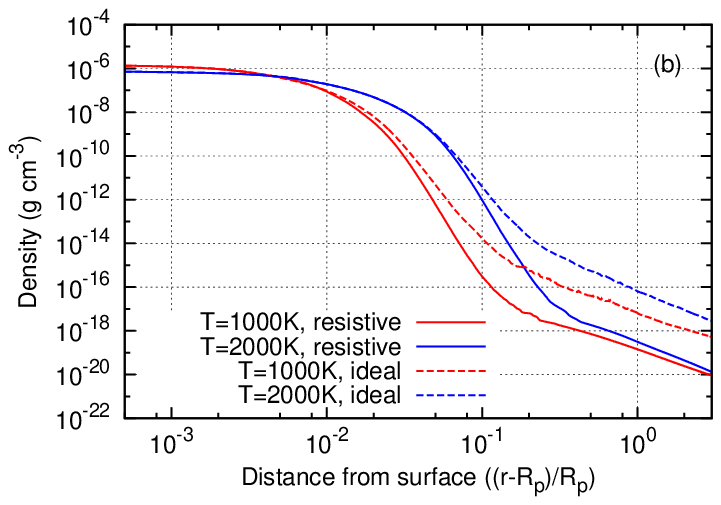}
\includegraphics[]{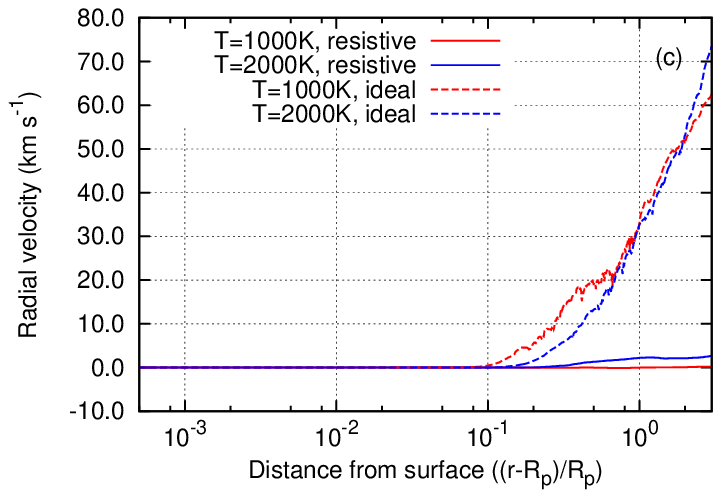}
\caption{
Comparison of the atmospheric structure between ideal and resistive MHD calculations.
(a) Temperature structure, (b) density profile, and (c) radial velocity profile.
The solid and dashed lines correspond to resistive and ideal MHD cases, respectively. The red and blue lines correspond to the surface temperature, $T_0=$1000K and 2000K, respectively.
}
\label{comparison}
\end{figure}

Figure \ref{comparison} compares the time-averaged atmospheric structure of resistive MHD cases with that of ideal MHD cases.
Note that all atmospheric structures in this paper are time-averaged, and all properties have large time variability.
The gas density in the upper atmosphere decreases more gradually in the ideal MHD cases because a larger amount of Alfvenic wave energy can reach the upper atmosphere and lifts up the gas by the magnetic pressure associated with the waves.
On the other hand, in the resistive MHD cases \Alfvenic waves are damped in the cool atmosphere at low altitudes because of the high resistivity, and then, the gas density sharply drops without the support from the magnetic pressure.
The \Alfvenic waves that enter the upper atmosphere contribute to the acceleration of planetary winds. The wind speeds are quite slow in the resistive MHD cases, while the wind are accelerated to several tens of kilometer per second in the ideal MHD cases. 
As a result, the mass-loss rates in the resistive MHD cases are much smaller than those in the ideal MHD cases, because the density in the upper atmosphere is around two orders of magnitude less than that of the ideal MHD cases, and the velocity is also much smaller.
If a planet is isolated or have a large distance from the central star, the planetary wind does not stream out in the resistive MHD condition except for intermittently driven outflows observed in the simulations because the average velocity is smaller than the escape velocity.
However, if the planet is a hot Jupiter having small semi-major axis, the gas overflows easily from the small Roche lobe and is blown away by the stellar wind from the central star.

In contrast to the large difference in the $\rho$ and $v_r$ profiles, the temperature structures of the resistive and ideal MHD cases are not so different each other.
Although the locations of the temperature inversion are slightly different, the upper atmospheres are heated up to $\sim 10^{4}\,{\rm K}$ in both ideal and resistive MHD cases.
Therefore, the heating by the dissipation of MHD waves is still important in the atmospheres of the resistive MHD cases.


\subsection{Dependence on Surface Temperature}\label{dependence_on_surface_temperature}


We show the relation between the surface temperature, ${T_0}$, of gaseous planets and the atmospheric structure, and then we discuss the dependence of the mass-loss rate.
In this subsection we fix the radius to the Jupiter radius and the amplitude of the velocity dispersion to 20\% of the sound speed at the surface.
Therefore, larger $T_0$ corresponds to larger MHD wave energy that is deposited to the magnetic flux tube.
$T_0$ is thought to be mainly determined by the irradiation from the central star, more specificly, determined by the combination of the effective temperature of the central star and the semi-major axis of the planet.
This implies that, in the framework of the wave-driven planetary winds, when we take planetary systems that have a similar central star, properties of the mass loss from hot Jupiters strongly depend on the distance from a central star.


\subsubsection{Atmospheric Structure}\label{atmospheric_structure}


\begin{figure}[h]
\includegraphics[]{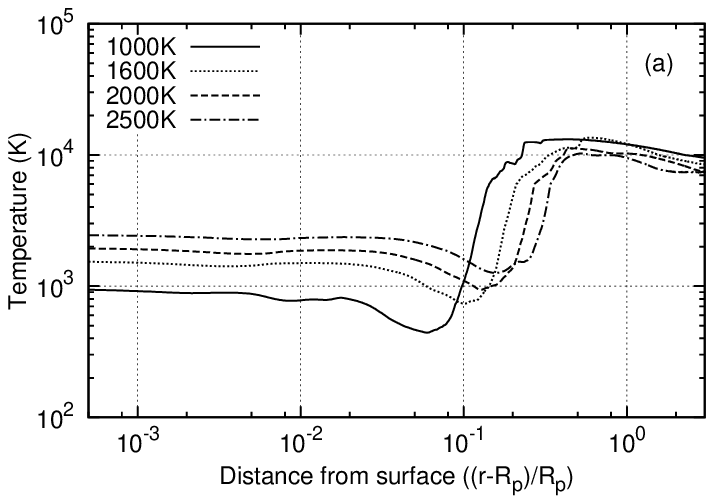}
\includegraphics[]{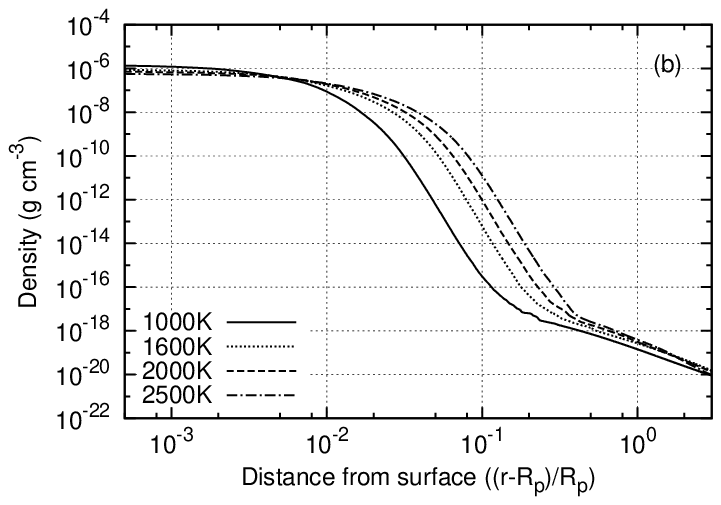}
\includegraphics[]{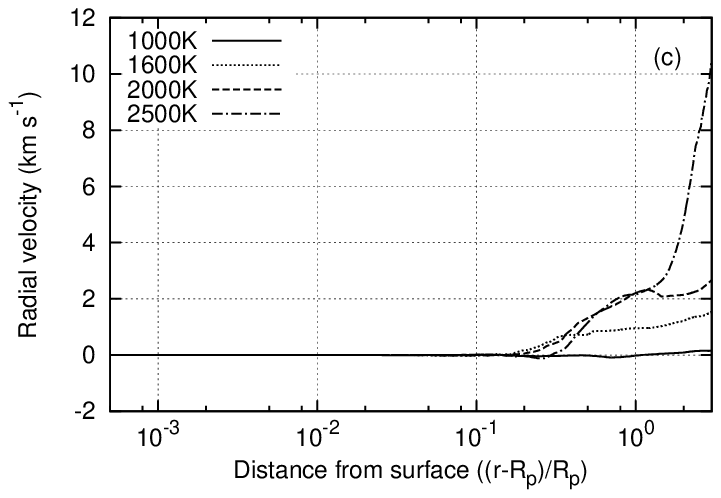}
\caption{
$T_{0}$ dependence of the atmospheric structure.
(a) Temperature structure, (b) density profile, and (c) radial velocity profile.
The horizontal axis denotes the distance from the planetary surface that is normalized by $R_{p}$ in logarithmic scale.
The solid, dotted, dashed and dot-dashed lines correspond to $T_{0}=$ 1000 K, 1600 K, 2000 K, and 2500 K, respectively.
}
\label{temp-temp}
\end{figure}

The dependence of the atmospheric structure on $T_{0}$ is shown in Figure \ref{temp-temp}.
In these simulations, the radius and mass of the planets are set to the radius and mass of Jupiter.
In all cases, the gas temperatures in the lower atmosphere are almost isothermal, and they rise rapidly in the upper atmosphere because of the dissipation of the Poynting flux carried by \Alfvenic waves.
The dissipation of the energy in the upper atmosphere makes a high temperature ($\sim$10000 K) corona-like region around the hot Jupiters.
In all the cases, a low-temperature region forms just below the location of the temperature inversion because of the adiabatic expansion of the gas.
As shown in Figure \ref{temp-temp} (a), the altitude of the temperature inversion is higher for higher $T_{0}$.
This is mainly caused by the difference of the density in the upper atmosphere.
In cases with higher $T_0$, the density in the upper atmosphere is higher because of the larger scale height; the density decreases more gradually with altitude (Equation (\ref{densitystructure})).
Since larger heating is required to heat up the denser gas, the temperature inversion is located at a higher altitude in cases with larger $T_{0}$. 

While the temperature and density panels in Figure \ref{temp-temp} show more or less similar profiles each other, there is a large difference in the radial velocity profiles.
The cases with low surface temperature $T_{0}\le 2000$ K yield quite slow wind velocities $< 3$ km s$^{-1}$.
In contrast, in the case with high $T_{0}$ of 2500 K, gas is largely accelerated to attain much faster velocity, $\approx 10$ km s$^{-1}$, than the other three cases.
These results of the velocity and density profiles suggest that the wind structure can be categorized into two regimes; while slow and weak planetary winds stream out from cooler planets, fast and strong planetary winds are accelerated from hotter planets.
We discuss this difference in Section \ref{two_regimes_of_the_Poynting_flux}.

\begin{figure}[h]
\includegraphics[]{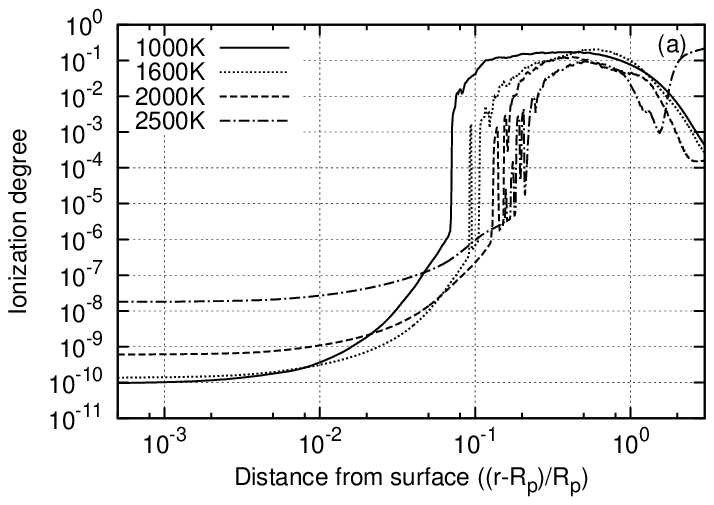}
\includegraphics[]{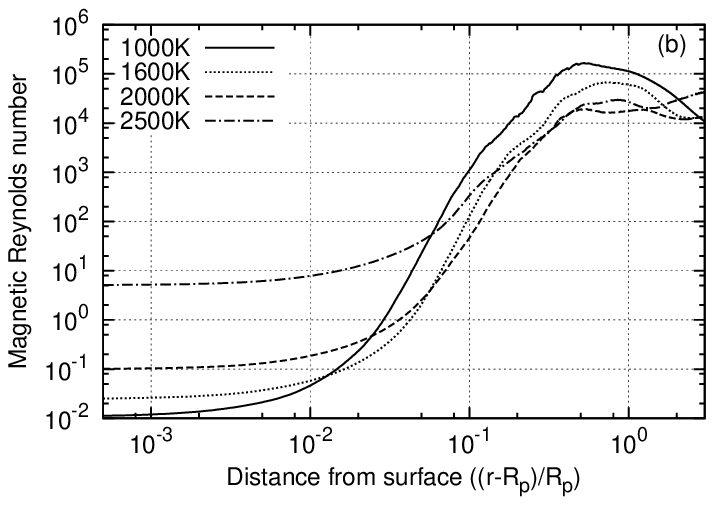}
\caption{
$T_{0}$ dependence of the (a) ionization degree and (b) magnetic Reynolds number in the atmosphere.
The horizontal axis denotes the distance from the planetary surface that is normalized by the $R_{p}$ in logarithmic scale.
The solid, dotted, dashed and dot-dashed lines correspond to $T_{0}=$ 1000 K, 1600 K, 2000 K, and 2500 K, respectively.
}
\label{temp-ionresis}
\end{figure}

Next, we examine the ionization degree that determines the magnetic resistivity in the atmosphere.
To evaluate the non-ideal effects in the weakly-ionized atmosphere, we introduce a magnetic Reynold number,
\begin{eqnarray}
R_{m}=\frac{vL}{\eta}
\end{eqnarray}
where $v$ and $L$ are a typical velocity and length in the system.
We use the scale height of the atmosphere (Equation (\ref{scaleheight})) and the amplitude of the velocity dispersion at the surface for $L$ and $v$, respectively.
Figure \ref{temp-ionresis} shows the dependence of $x_e$ and $R_m$ on $T_0$.
Since the typical value of $T_{0}$ is too low to ionize hydrogen atoms, the ionization degree is very low in the lower atmosphere.
$x_e\sim10^{-10}$ in the cases with $T_0=$ 1000 K and 1600K, and {$x_e$ is still $\sim10^{-8}$} even for $T_0=$ 2500 K.
In this temperature region, the ionization degree is supported by the ionization of alkali metals, such as sodium and potassium, that have relatively low ionization potential.
This low $x_e$ leads to high magnetic resistivity and small magnetic Reynolds number.
As a result, the magnetic Reynolds number in the lower atmosphere is very small, $R_m\sim 10^{-2}$ in the cases with $T_0=$ 1000 K and 1600K, and still $R_m\sim 1$ for $T_0=$ 2500 K, although it is $\approx$100 times larger owing to the higher $x_e$.
The resultant low $R_{m}$ means that the non-ideal effects are important in the lower atmosphere.
In the upper atmosphere, however, high gas temperature and low gas density cause a sudden surge of $x_e$.
$x_e$ increases gradually in the region below the temperature inversion region as $\rho$ decreases with altitude.
$x_e$ is very high above there because the temperature rises high enough to ionize the hydrogen atoms.


\subsubsection{Two Regimes of the Poynting Flux}\label{two_regimes_of_the_Poynting_flux}

\begin{figure}[h]
\includegraphics[]{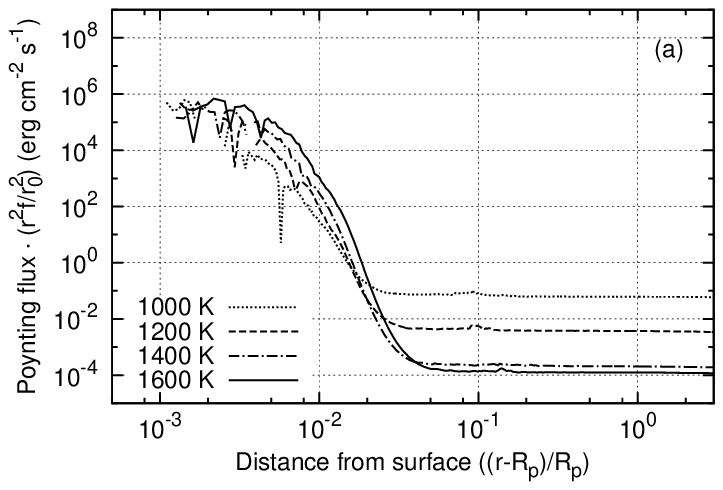}
\includegraphics[]{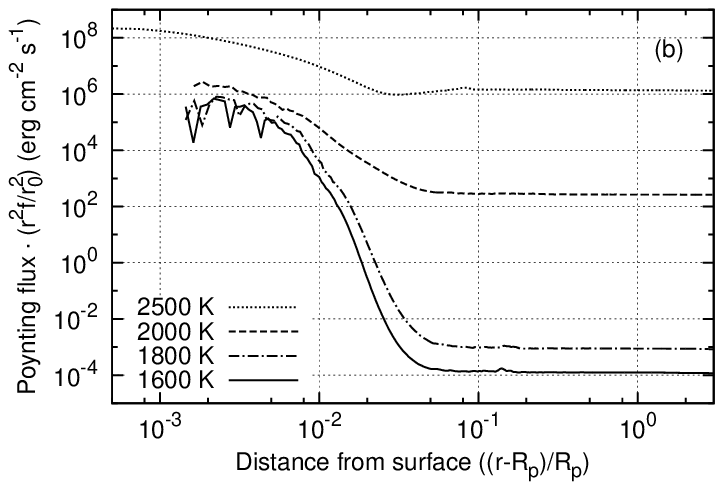}
\caption{
Net outgoing Poynting flux as a function of the distance from the surface for different $T_0$ cases.
(a) presents cases with lower $T_0\le 1600$ K and (b) presents cases with higher $T_0\ge 1600$ K.
The Poynting flux is multiplied by $r^{2}f\!\left(r\right)/r^{2}_{0}$ to separate the effect of the adiabatic expansion of the flux tube.
}
\label{temp-poynting}
\end{figure}

Figure \ref{temp-poynting} shows the net Poynting flux in the atmosphere for cases with different $T_{0}$.
The net outgoing Poynting flux decreases rapidly with an increase of the altitude in the lower atmosphere, and it becomes almost constant in the upper atmosphere.
The Poynting flux injected at the surface increases monotonically with $T_0$, but its behavior in the atmosphere is not simple; the net Poynting flux in the upper atmosphere is smaller for higher $T_0$ when $T_{0}\le 1600\,{\rm K}$ (a), but it is larger for higher $T_0$ when $T_0>1600$K (b).
The \Alfvenic waves are damped in the lower atmosphere where the magnetic resistivity is high.
The scale height of the atmosphere plays a key role in the resistive dissipation of the \Alfvenic waves.

The scale height is larger for higher $T_0$, which gives the large density in the upper atmosphere.
On the one hand the larger density in the upper atmosphere leads to a larger amount of the mass loss, but on the other hand it gives a negative effect on the wind through the enhancement of the resistive dissipation of the \Alfvenic waves.
The ionization degree is lower in the higher-density gas because of the efficient recombination, and consequently the magnetic resistivity is relatively higher.
In Figure \ref{temp-ionresis} (b), the magnetic Reynolds number of the case with $T_0 = 1000$ K is low in the lower atmosphere, but it increases quickly in the upper atmosphere.
In the case with $T_0 = 1600$ K, it is as low as the $T_0=$1000 K case in the lower atmosphere, but it increases more slowly than in the 1000 K case. Therefore, the magnetic Reynolds number remains relatively lower in the almost entire region than in the $T_0=$1000 K case.
Because of the larger resistivity, a smaller fraction of the injected wave energy reaches the upper atmosphere in the $T_0=1600$ K case, compared to the $T_0=1000$ K case.
This is the main reason why the transmitted Poynting flux shows the negative tendency on the surface temperature for $T_0 < 1600$ K.
However, for $T_0 >1600$ K, the Poynting flux exhibits a positive correlation with $T_0$ because the magnetic resistivity is lower for higher $T_0$.

\begin{figure}[h]
\includegraphics[]{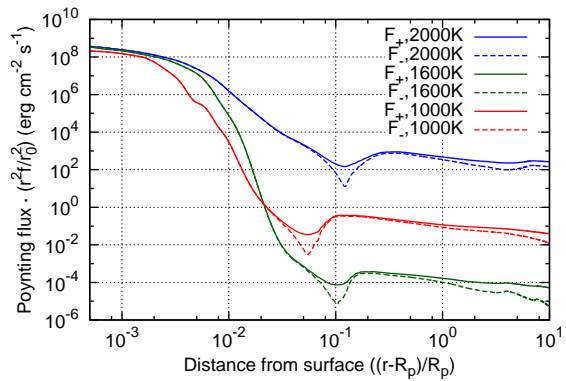}
\caption{
Comparison of the inward and outward Poynting flux on distance from the surface for the cases with $T_{0}=$ 1000 K (red), 1600 K (green), and 2000 K (blue).
The Poynting flux is multiplied by $r^{2}f\!\left(r\right)/r^{2}_{0}$ in order to remove the effect of the adiabatic expansion of the flux tube.
The solid lines are the outward energy flux $F_{+}$, and the dashed lines are the inward energy flux $F_{-}$.
}
\label{inandout}
\end{figure}

Figure \ref{inandout} compares the outward and inward energy flux the \Alfvenic waves (equation \ref{Pflux2}).
Here we show three cases with $T_{0}$ of 1000 K, 1600 K, and 2000 K.
Every case shows that the inward energy flux tracks the outward energy flux with a slightly smaller level in the almost entire region of the atmosphere.
The net outgoing Poynting flux in Figure \ref{temp-poynting} is considerably smaller than the outward energy flux in Figure \ref{inandout}. 
These results indicate that the injected outgoing \Alfven waves are reflected back downward through the propagation in the atmosphere.
A tiny fraction of the injected \Alfven waves can reach the upper atmosphere, which contributes to the heating and the acceleration of the gas.
In general, a small fraction of the injected Poynting flux by \Alfven waves is transmitted to the upper atmosphere because they suffer dissipation and reflection, whereas the resistivity is one of the keys that control the dissipation of the waves.
The fraction that enters the upper atmosphere determines the properties of the wind.
For instance, if a large amount of the MHD wave energy reaches the upper atmosphere, it will rise the density  there and drive outflows.

The effect of the resistive dissipation of the \Alfvenic waves is also seen in Figure \ref{inandout}.
Both dissipation and reflection of the waves decrease the Poynting flux.
The Poynting flux in the case $T_0=$1000 K (red lines) decreases much faster than that of the case $T_0=$2000 K (blue lines), which is caused by both resistive dissipation and reflection.
The higher magnetic resistivity in the 1000 K case makes the Alfvenic waves dissipated faster, and the steeper density gradient results in strong reflection.
In the 1600 K case (green lines), the Poynting flux quickly drops, which is much faster than in the other two cases.
Since the scale height is larger than that of the $T_0=$1000 K case, the reflection of the \Alfvenic waves is suppressed.
In this case the resistive dissipation of the \Alfvenic waves is the primary reason of the rapid drop of the Poynting flux, and the amount of the energy that can reach the upper atmosphere is quite small.


\subsubsection{Time Variability of Mass-loss Rate}\label{time_variability_of_mass-loss_rate}

\begin{figure}[h]
\includegraphics[]{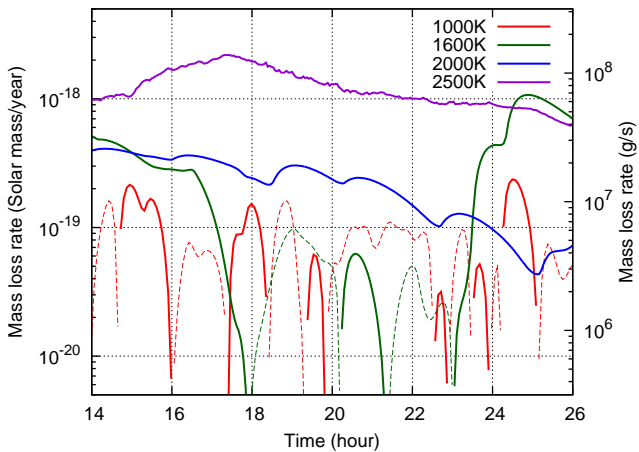}
\caption{
Time evolution of the mass-loss rates for the cases with $T_0=$ 1000 K 
(red), 1600 K (green), 2000 K (blue), and 2500 K (violet).
Thick solid lines show the outflow rates, and dashed lines show the infall rates.
On the left the mass-loss rate is shown in units of solar mass per year and on the right in units of gram per second.
}
\label{temp-mslsvar}
\end{figure}

Our simulations show large time variabilities of the atmospheric structure.
Figure \ref{temp-mslsvar} shows the mass-loss rate for different $T_0$ cases with time. 
One can see that in the cases with lower $T_0\le 1600$ K the outflow is not time-steady, but the atmospheric material infalls during certain periods, which is seen as discontinuous lines.
For example, in the case with $T_{0}=$ 1000 K, the outflow phases are relatively shorter and it is comparable to the total duration of the infall phases.
In the case with $T_{0}=$ 1600 K, the integrated duration of the outflow phases is longer, and the infall phase disappears for higher $T_{0}\ge 2000$ K.

These results suggest that in the cases with low $T_0\le 1600$ K the atmosphere is almost hydrostatic because the damping of \Alfvenic waves is too severe to drive steady wind, and the outflows occur only intermittently.
For higher $T_0$ the intermittency of the planetary winds disappears, more steady outflows are obtained.

\begin{figure}[h]
\includegraphics[]{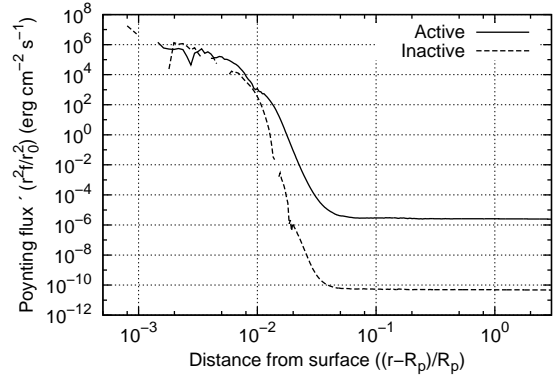}
\caption{
Comparison of the Poynting flux in the outflow (solid) and infall (dashed) phases for the case with 
$T_{0}=$ 1600K.
The Poynting flux is multiplied by $r^{2}f\!\left(r\right)/r^{2}_{0}$ in order to remove the effect of the adiabatic expansion of the flux tube.
}
\label{pflux1600comp}
\end{figure}

In Figure \ref{pflux1600comp} we compare the Poynting flux by the \Alfven waves in the active phase during 23.6-31.6 hour and that in the infall phase during 17.8-20.2 hour in Figure \ref{temp-mslsvar}.
In the outflow phase, a larger amount of the Poynting flux is carried by the \Alfven waves into the upper atmosphere, while in the infall phase, the Poynting flux in the upper atmosphere is strongly reduced around four orders of magnitude.
This result implies that the transmission of the Poynting flux into the upper layor controls the on-off nature of the atmospheric escape for the low $T_0$ cases.
However, we would like to note that the case with $T_0=1000$ K (not shown) gives 
more complicated behavior; Poynting flux during outflow phases does not always exceed Poynting flux during infall phases bucause of the rapid temporal variations.


\subsection{Dependence on Planet Radius}\label{dependence_on_planet_radius}


Here we describe the dependence of the atmospheric structure and the mass-loss rate on the radius of gaseous planets $R_{p}$.
A number of theoretical works on the thermal evolution of the gaseous planets suggest that radii of gaseous planets, which tend to be around Jupiter radius, are affected by surrounding circumstances \citep[e.g.,][]{bur07a}.
In addition, observations have shown that some hot Jupiters have enormously larger radii than that are expected \citep{bar10}.
Therefore, we treat $R_{p}$ as a parameter and investigate the dependence on $R_{p}$.
The mass-loss rate is very small for low $T_0$, therefore we set $T_0=$2000 K in this subsection.
We also fix the amplitude of the velocity dispersion at the surface, $\delta v=0.2 c_s$.

\begin{figure}[h]
\includegraphics[]{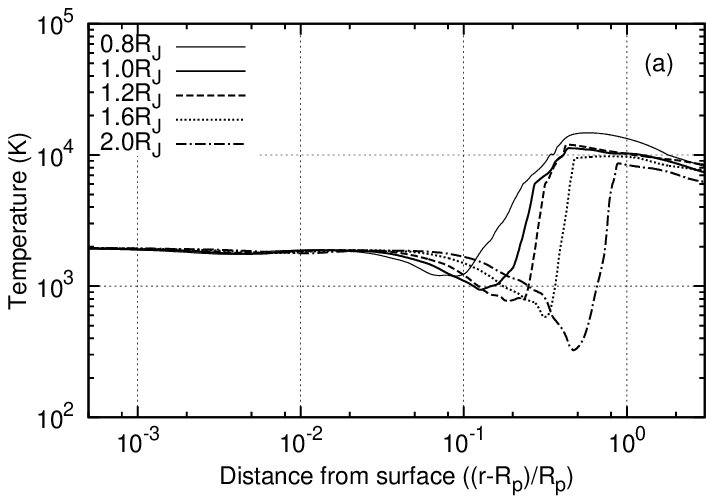}
\includegraphics[]{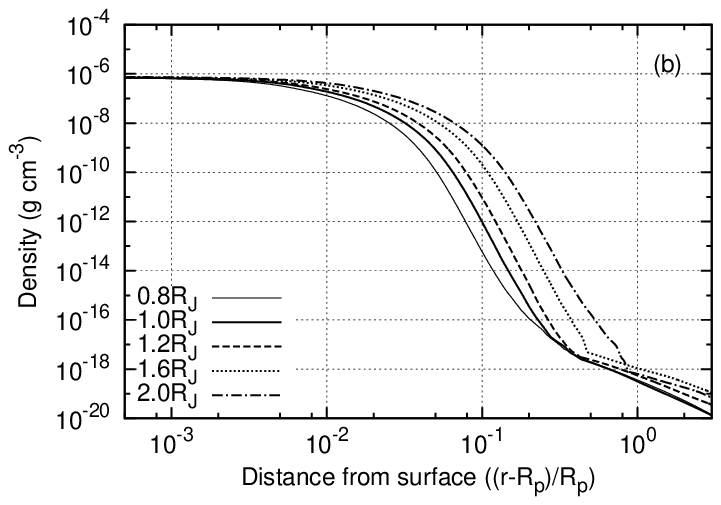}
\includegraphics[]{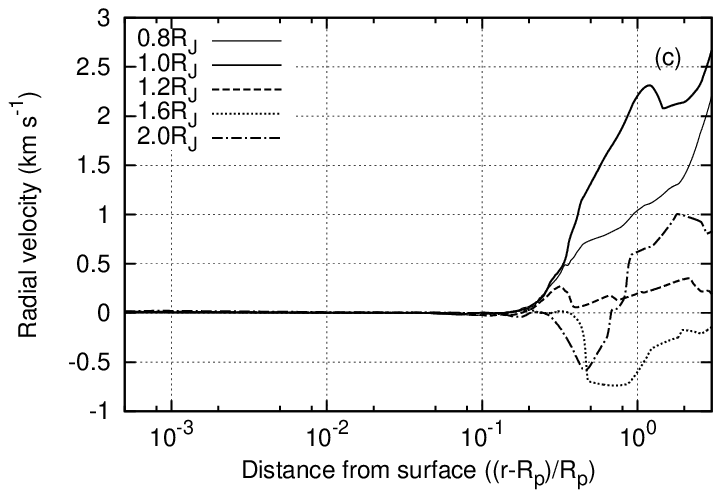}
\caption{
$R_{p}$ dependence of the atmospheric structure.
(a) Temperature structure, (b) density profile, and (c) radial velocity profile.
The narrow solid, solid, dashed, dotted, and dot-dashed lines correspond to $R_{p}=$ 0.8$R_{J}$, 1.0$R_{J}$, 1.2$R_{J}$, 1.6$R_{J}$, and 2.0$R_{J}$, respectively.
Distance from the surface (horizontal axis) is normalized by $R_{p}$ each case.
}
\label{radius-temp}
\end{figure}

Figure \ref{radius-temp} shows the dependence of the atmospheric structures on $R_p$.
At low altitudes the temperature is almost constant ($\approx$ surface temperature) in all the cases. It decreases initially but rises eventually by the heating owing to the wave dissipation.
The temperature inversion is located at a higher altitude for larger $R_{p}$, because the scale height is larger; the slower decrease of the density leads to higher density in the upper region (Figure \ref{radius-temp}), which is not heated up to high temperatures.
Accordingly, the bending location of the density gradient, which corresponds to the location of the temperature inversion, is systematically higher for larger $R_{p}$ by the same reason.

In all the cases, the temperature of the upper atmosphere is elevated to $\sim10^{4}$ K, by the dissipation of MHD waves.
However, the wind velocity is quite slow.
Particularly in the cases with the large $R_{p}$ $\ge 1.2R_{J}$, the gas in the atmosphere is almost static or even infalls partly, whereas the cases with the smaller radius show outflows with a few km s$^{-1}$.

\begin{figure}[h]
\includegraphics[]{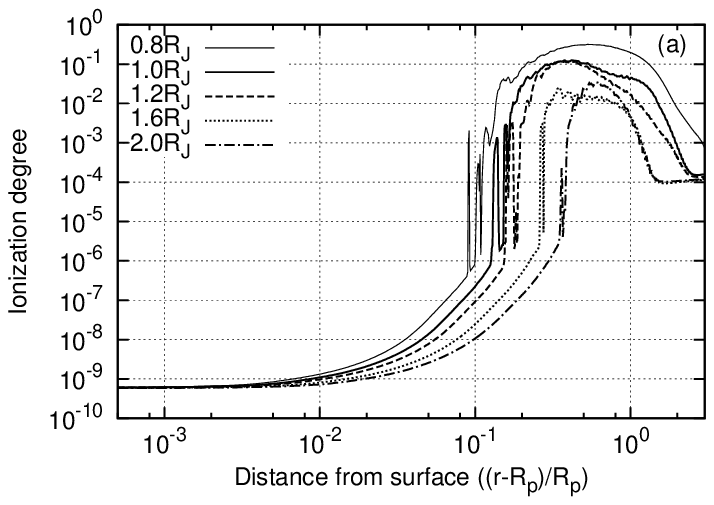}
\includegraphics[]{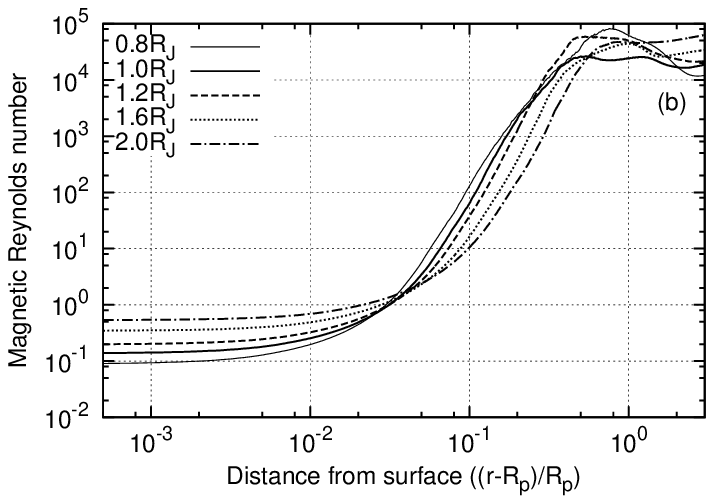}
\caption{
$R_{p}$ dependence of the (a) ionization degree and (b) magnetic Reynolds number in the atmosphere.
The narrow solid, solid, dashed, dotted, and dot-dashed lines correspond to $R_{p}=$ 0.8$R_{J}$, 1.0$R_{J}$, 1.2$R_{J}$, 1.6$R_{J}$, and 2.0$R_{J}$, respectively.
Distance from the surface (horizontal axis) is normalized by $R_{p}$ each case.
}
\label{radius-ionresis}
\end{figure}

Figure \ref{radius-ionresis} shows the $R_{p}$ dependence of the ionization degree and magnetic Reynolds number.
The ionization degree, $x_e$, is quite low at low altitudes because of the low temperature there ($\approx 2000$ K) in all the cases.
The dominant ion sources in the lower atmosphere are the alkali metals.
In contrast, $x_e$ increases to $10^{-4}-10^{-1}$ in the upper atmosphere in all cases.
In the smaller $R_{p}$ case, $x_e$ starts to increase from a lower altitude and the final $x_e$ is also higher in the upper atmosphere, because the recombination is slower there due to the lower density.

The mass-loss rate remains very small in these cases.
The case with $R_p=R_{J}$ gives the mass-loss rate of $2.6\times10^{-19}\,M_{\odot}\,\rm{yr^{-1}}$, and the case with 0.8 $R_{J}$ gives $2.3\times10^{-19}\,M_{\odot}\,\rm{yr^{-1}}$, that correspond to $1.7\times10^{7}\,{\rm g\,s^{-1}}$, and $1.4\times10^{7}\,{\rm g\,s^{-1}}$, respectively (here, $M_p=M_J$ and $T_0=2000$ K).
In the ideal MHD calculations, the mass-loss rate in the case with $R_{p}=R_{J}$ and $T_{0}=2000\,{\rm K}$ is $\sim1.9\times10^{10}\,{\rm g\,s^{-1}}$\citep{tan14}.
Our results show that the large resistivity as a result of the low ionization degree in the low atmosphere significantly suppresses the mass loss from hot Jupiters.
However, we would like to note that the wind shows large fluctuations.
There is a possibility remained that the planetary wind streams out in an intermittent manner.
Also, the Roche lobe of a hot Jupiters is small because of the small separation from the central star.
Then, once a slow gas material overflows the Roche lobe, it will be blown away by the stellar wind from the central stars.


\subsection{Dependence on Velocity Dispersion}\label{dependence_on_velocity_dispersion}


\begin{figure}[h]
\includegraphics[]{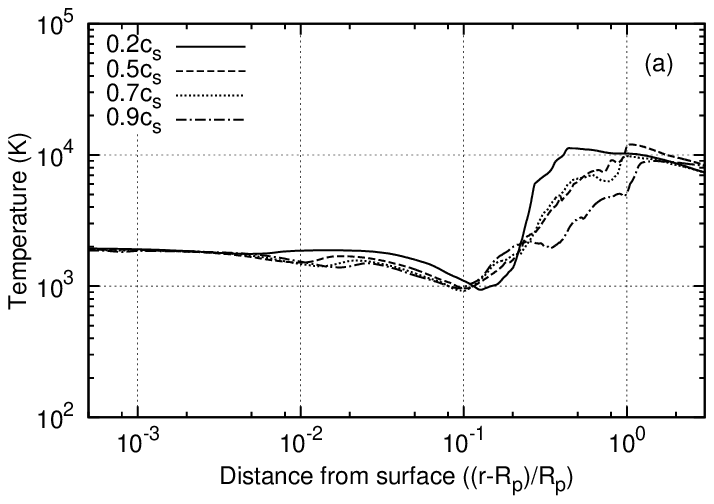}
\includegraphics[]{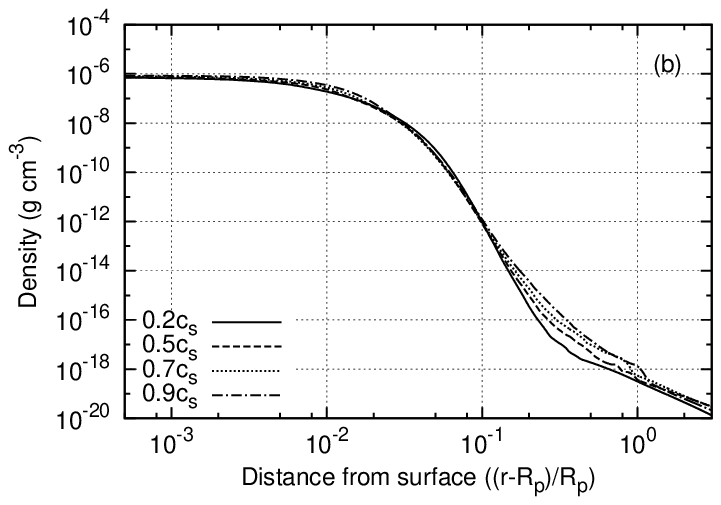}
\includegraphics[]{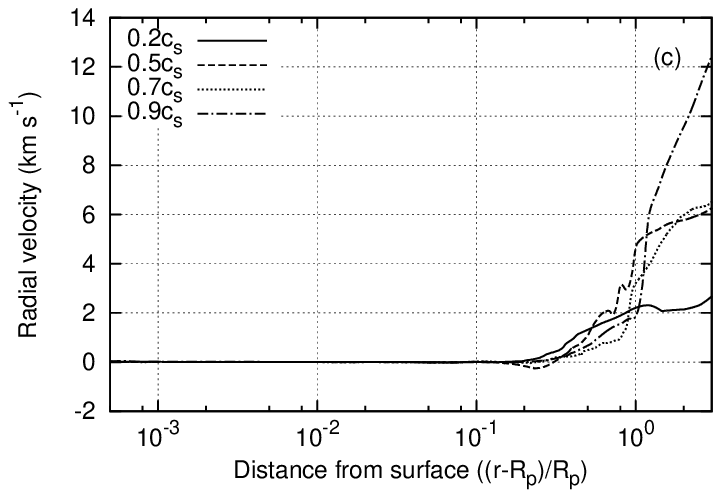}
\caption{
Dependence of the atmospheric structure on the $\delta v$ at the surface.
(a) Temperature structure, (b) density profile, and (c) radial velocity profile. The horizontal axis denotes the distance from the planetary surface that is normalized by the $R_{p}$ in logarithmic scale.
The solid, dashed, dotted, and dot-dashed lines correspond to $\delta v=$ $0.2c_{s}$, $0.5c_{s}$, $0.7c_{s}$, and $0.9c_{s}$, respectively, and $c_{s}$ is the sound speed at the surface.
In all cases, $R_{p}=R_{J}$, $M_{p}=M_{J}$, and $T_{0}=2000\,{\rm K}$.
}
\label{dv-temp}
\end{figure}

The velocity amplitude at the surface, $\delta v$, is also an important parameter that controls the structure of the atmosphere and the wind of a planet.
In this subsection, we fix $R_{p}=R_{J}$, $M_{p}=M_{J}$, and $T_{0}=2000\,{\rm K}$.
In Figure \ref{dv-temp}, we show the atmospheric structure of the cases with different $\delta v$.
Cases with different $\delta v$ give similar density and temperature profiles.
On the other hand, larger $\delta v$ gives faster wind velocity.
Because of the similarity of the density and the temperature shown in Figure \ref{dv-temp}, the ionization degree and the magnetic resistivity are also similar in all the cases.
Therefore, it can be said that the difference of $\delta v$ at the surface has small effect on the structure of the atmosphere except for the radial velocity.

\begin{figure}[h]
\includegraphics[]{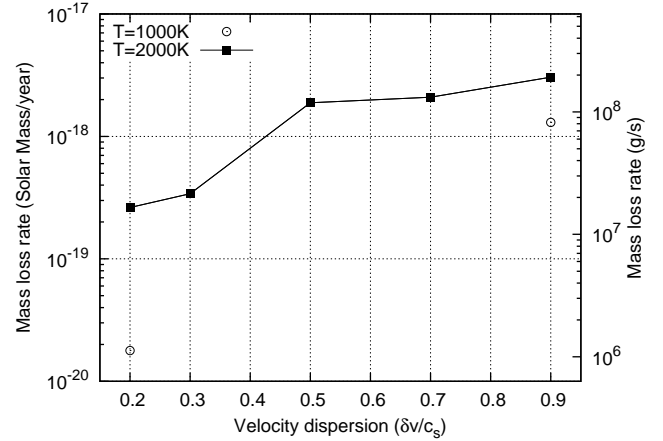}
\caption{
Relation between $\delta v/ c_s$ and the mass-loss rate for $T_0=1000$ K (circle) and 2000 K (filled square).
The mass-loss rate is shown in units of solar mass per year (left axis) and gram per second (right axis).
}
\label{dv2-massloss}
\end{figure}

Though the difference of the density in the upper atmosphere is very small, the wind velocity is faster for larger $\delta v$, and it causes the increase of the mass-loss rate as shown in Figure \ref{dv2-massloss} for cases with $T_0=1000$ K and 2000 K.


\section{Discussion}\label{discussion}



\subsection{Ohmic Heating in Planetary Atmosphere}\label{ohmic_heating_in_planetary_atmosphere}


We have investigated the effect of the magnetic diffusion on the dissipation of the \Alfvenic waves, which suppresses the planetary winds, so far. 
On the other hand, the resistive dissipation also heats up the ambient gas via Ohmic (Joule) heating, which plays a positive role in driving outflows. In this subsection, we examine the role of the Ohmic heating in a quantitative manner.

Current density is
\begin{eqnarray}
{\mbf j}=\frac{c}{4\pi}\left(\nabla\times{\mbf B}\right),
\end{eqnarray}
and electric conductivity $\sigma$ is related with the magnetic resistivity $\eta$ by
\begin{eqnarray}
\frac{1}{\sigma}=\frac{4\pi}{c^{2}}\eta.
\end{eqnarray}
Then, Ohmic heating rate per unit volume, $Q_{\rm Ohm}={\mbf j^{2}}/\sigma$, is derived as
\begin{eqnarray}
Q_{\rm Ohm}=\frac{\eta}{4\pi}\left(\nabla\times{\mbf B}\right)^{2}.
\end{eqnarray}

\begin{figure}[h]
\includegraphics[]{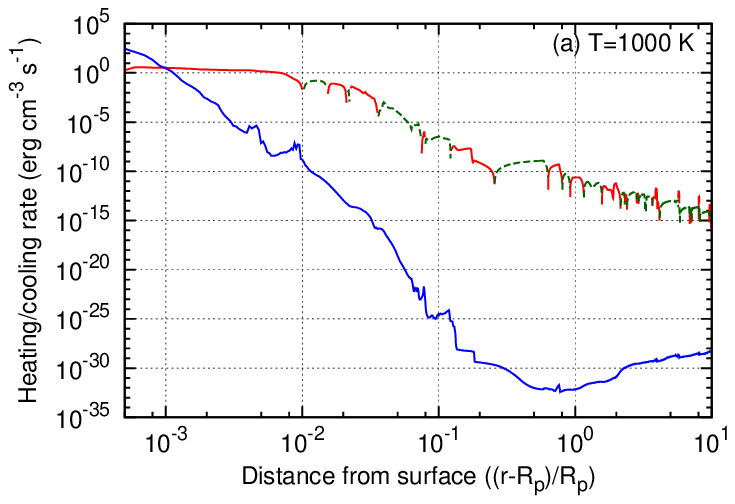}
\includegraphics[]{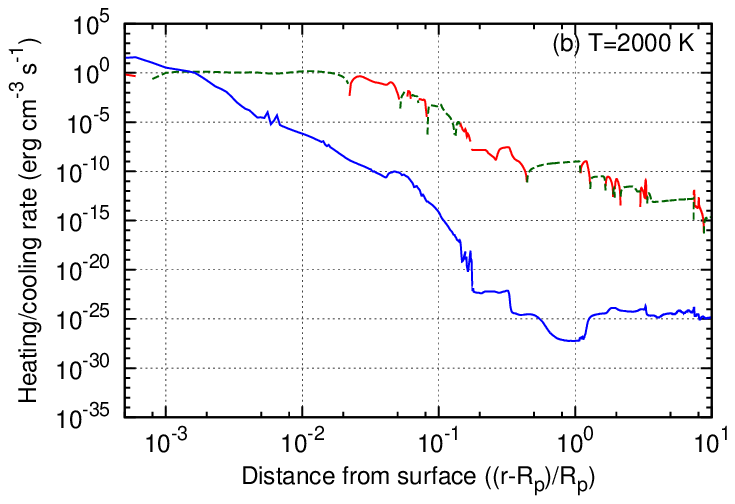}
\caption{
Comparison of ideal MHD heating rate (red solid), cooling rate (green dashed), and Ohmic heating rate (blue solid) per unit volume {for cases with 
$T_0=1000$ K (a) and 2000 K (b)}.
We set $\delta v =0.2c_s$, $M_p = M_J$, and $R_p = R_j$.
}
\label{ohmiccomp}
\end{figure}

Figure \ref{ohmiccomp} shows the Ohmic heating rate in comparison with the ideal MHD heating (red solid) and cooling (green dashed) rates of the gas. 
The ideal MHD heating includes shock heating (see Section \ref{Poynting_flux}) in addition to adiabatic heating.
In the lower atmosphere, the Ohmic heating dominates over, or be comparable to the ideal MHD heating because of the large resistivity.
However, since the \Alfvenic waves dissipate very rapidly, the Ohmic heating rate drops drastically as the altitude increases.

\begin{figure}[h]
\includegraphics[]{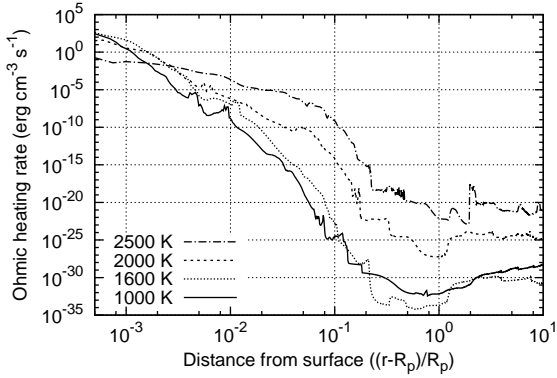}
\caption{
Dependence of the Ohmic heating rate per unit volume on $T_{0}$.
The solid, dotted, dashed, and dot-dashed lines correspond to $T_{0}=$1000 K, 1600 K, 2000 K, and 2500 K, respectively.
}
\label{ohmic}
\end{figure}

Resistivity depends sensitively on temperature.
Figure \ref{ohmic} compares the Ohmic heating rates for cases with different $T_0$.
The higher $T_{0}$ cases have smaller Ohmic heating rate in the lower atmosphere, but it reverses in the upper atmosphere, because the dissipation of the \Alfvenic waves is more gentle at low altitudes in these cases and a larger amount of the wave energy reaches the upper atmosphere.
In addition, as shown in the Figure \ref{temp-ionresis}, the magnetic Reynolds number is relatively lower in the upper atmosphere for higher $T_0$, which further increases the resistive heating.

However, the Ohmic heating rate is quite small in comparison to the shock heating except for the region closed to the surface (Fig. \ref{ohmiccomp}), so this effect is negligible in the upper atmosphere and gives little contribution to the atmospheric escape.
Even in the low atmosphere where the Ohmic heating is large, it does not play a primary role.
The comparison between the ideal and non-ideal calculations in Figure \ref{comparison} (a) shows that the temperature profiles in the lower atmosphere are almost same.
In fact, the temperatures in the lower atmosphere in the resistive cases are barely higher than the ideal cases and the difference is only several tens of kelvin.
In summary, the Ohmic heating does not play important roles in the entire structure of the atmosphere and the atmospheric escape.


\subsection{A Possibility for Bipolar Outflows}\label{geometry_of_magnetic_flux_tube}


In this work and the previous work, we have fixed the shape of the open magnetic flux tube that is described in Section \ref{numerical_method}.
We presume that the overall structure of magnetic field lines of gaseous planets is similar to that of the sun.
The sun has a multipole magnetic field and the properties of open magnetic flux tubes are observed in detail \citep[e.g.,][]{tsu08,ito10}.
However, the actual structures of magnetic field in exoplanets are unknown.
The overall structure of the planetary magnetic field affects the shape of the open magnetic flux tube.
The most critically affected parameter in our modeling is the filling factor $f\!\left(r\right)$ (see Equation (1) and (2) of \cite{tan14}).  
If the planetary magnetic field has a simple dipole structure, open magnetic flux tubes appear in the magnetic polar regions.
In this case, heating in the upper atmosphere and the acceleration of gas flow by MHD wave energy occur only in the polar regions.
Therefore, the actual appearance of flow is expected to be bipolar.
In the simple dipole case of the magnetic field, the degree of super-radial expansion of the open magnetic flux tube should be smaller, and thus the mass-loss rate is expected to be smaller because the reflection of the energy flux is expected to be enhanced.
To investigate this more quantitatively, a three-dimensional calculation is required, which remains difficult at present.
On the other hand, the observation of the bipolar outflow from an exoplanet is, in principle, possible in the detailed analysis of the differential spectroscopy for transiting hot Jupiters, unless the magnetic dipole axis is perpendicular to both the orbital plane of the exoplanet and the line-of-sight.
This is because a blue-shifted outflow before ingress and a red-shifted outflow after the emersion, or vice versa, can be seen in this case.
Once the mass flux of the bipolar outflow from the exoplanet is observed, we will possibly constrain the property of magnetic field of the exoplanet by assuming the driving mechanism as described in this work.


\subsection{Limitations in the Present Model}\label{limitation}


As we mentioned in the introduction, high energy radiation such as X-ray and extreme ultraviolet from a central star is important for the heating of upper atmosphere and atmospheric escape from hot Jupiters.
Here we compare the height where the heating by MHD waves is important and the penetration depth of stellar XUV.
It is useful to introduce $R_{\rm XUV}$, the radius at which XUV is absorbed.
$R_{\rm XUV}$ means that the optical depth for XUV reaches unity at that radius, and it corresponds to the radius where the column density of hydrogen becomes $\sim 5\times10^{17}\,{\rm cm^{-2}}$\citep[e.g.,][]{mur09,rog11}.
We calculate $R_{\rm XUV}$ in each calculations and compare it with the region where the heating by the dissipation of the MHD waves becomes important.

\begin{figure}[h]
\includegraphics[]{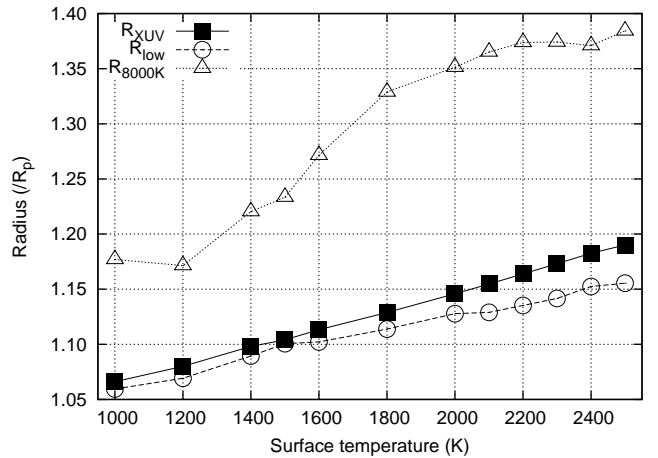}
\caption{
Dependence of $R_{\rm XUV}$ on $T_{0}$ and comparison with $R_{low}$ and $R_{8000{\rm K}}$.
The black squares with the solid line corresponds to $R_{\rm XUV}$, the circles with the dashed line and the triangles with the dotted lines correspond to $R_{low}$ and $R_{8000{\rm K}}$, respectively.
We set $\delta v=0.2c_{s}$, $R_{p}=R_{J}$, and $M_{p}=M_{J}$ for this comparison.
}
\label{rxuv}
\end{figure}

Figure \ref{rxuv} shows $T_{0}$ dependence of $R_{\rm XUV}$ and comparison with other considerable radii.
As shown in Figure \ref{temp-temp}(a), the gas temperature slightly drops due to adiabatic expansion and then heated drastically to $\sim10^{4}$ K.
$R_{low}$ is the radius at which the temperature become lowest and heating by MHD waves starts to become important, and $R_{8000{\rm K}}$ is the radius at which the temperature become higher than 8000 K.
In this range of $T_{0}$, $R_{\rm XUV}$ always locates only slightly above $R_{low}$ and the value varies between $1.05\sim1.20R_{p}$.
This is consistent with previous works on the XUV heating in the atmosphere of hot gaseous planets \citep[e.g.,][]{mur09}.
In addition, $R_{8000{\rm K}}$ always lies much higher altitude than $R_{\rm XUV}$.

These result suggest that the region heated to $\sim10^{4}$ K by the dissipation of the MHD waves is optically thin to XUV photons, and XUV can reach the altitude that the temperature becomes lowest in our model.
Photo-ionization, cooling from metal lines, Lyman $\alpha$ play important roles in the upper atmosphere of hot Jupiters.
At this stage, our model does not include the effects of stellar XUV irradiation, the inclusion of which should improve our model.
In the present modeling, the results cannot explain the observational results of the mass-loss rates adequately, since the mass-loss rates are reduced largely when we take into account the effects of the magnetic resistivity.
This seems to indicate that the resulting mass-loss rate strongly depends on the details of the input physics. More realistic calculation is critically needed, and will be out future work.


\section{Conclusion}\label{conclusion}


We have studied the structure of the atmosphere and the wind of hot Jupiters by non-ideal MHD simulations, particularly focusing on the effects of the magnetic diffusion.
The \Alfvenic waves generated from the surface are strongly damped via resistive dissipation in the weakly-ionized atmosphere at low altitudes and only a tiny fraction of the initial energy can reach the upper atmosphere. 
Consequently, the mass-loss rate is reduced significantly, compared to that obtained from the ideal MHD simulations.
However, the \Alfvenic waves that survive to transmit into the upper atmosphere still support the temperature inversion and heat up to $\sim\,10^{4}$ K by the wave dissipation; the \Alfvenic waves are still important in determining the temperature profile even though the magnetic diffusion is taken into account.
We would like to note that the wave heating here is not owing to Ohmic heating but the consequence of the shock dissipation of the compressive waves that are nonlinearly excited from the \Alfvenic waves.

We also investigated the dependence of atmospheric properties on the surface temperature, the planet radius, and the velocity dispersion at the surface.
We found a nonmonotonic dependence of the Poynting flux on the surface temperature (Figure \ref{temp-poynting}), because the resistivity, which controls the dissipation of the \Alfvenic waves, is determined globally the density and temperature of the atmosphere.
We point out that the heating and acceleration of gas will take place in magnetic polar regions if the planetary magnetic field is simple dipole structure, and non-spherically symmetric bipolar outflow will takes place.

In our calculations, the treatment of the radiative cooling is simplified especially for low temperature gas (see discussion part of \cite{tan14}).
Therefore, a more accurate treatment is needed for future investigation and now we are going to improve this part.

\section*{Acknowledgement}
We thank G. Harper for providing a useful code to calculate ionization degree in the gas.
We also thank the referee for valuable comments for improving the manuscript.
This work was supported in part by Grants-in-Aid for Scientific Research from the MEXT of Japan, 22864006 (TKS).

\end{document}